# Impurity potential fluctuations for selectively doped p-Ge/Ge$_{1-x}$Si$_x$ heterostructures in the quantum Hall regime

*Yu. G. Arapov[†], O. A. Kuznetsov[‡], V. N. Neverov[†], G. I. Harus[†], N. G. Shelushinina[†], M. V. Yakunin[†]*

[†]Institute of Metal Physics RAS Ural branch, Ekaterinburg GSP-170, 620219 Russia
[‡]Scientific Research Institute at Nizhnii Novgorod State University, Russia

**Abstract.** Two models for the long-range random impurity potential (the model with randomly distributed charged centers located within a layer and the model of the system with a spacer) are used for evaluation of the impurity potential fluctuation characteristics: the random potential amplitude, nonlinear screening length in vicinity of integer filling factors $\nu = 1$ and $\nu = 2$ and the background density of state (DOS). The described models are suitable for explanation of the unusually high value of DOS at $\nu = 1$ and $\nu = 2$, in contrast to the short-range impurity potential models.

The nature of the quantum Hall effect (QHE) is closely linked with a phenomenon of electron localization in a two-dimensional (2D) disorder system under quantizing magnetic field ($B$) [1,2]. The appearance of quantum plateaux in the $\rho_{xy}(B)$ dependences with vanishing values of $\rho_{xx}$ is now commonly accepted to be caused by the existence of disorder-induced mobility gaps in the density of states (DOS) of a 2D-system. When the Fermi level is settled down in the gap, the thermally activated behavior of $\rho_{xx}$ (or $\sigma_{xx}$) is observed due to excitation of electrons into very narrow bands of extended states centered at Landau level (LL) energies $E_N$. The DOS in mobility gaps may be evaluated from the data on activation energy $E_A$ as a function of the LL filling factor $\nu = n/n_B$ ($n$ is the electron density, $n_B = eB/hc$) [3-6]. The filling factor can be tuned by the change of either a carrier density [3] or a magnetic field [4-6].

We used the method of activated magnetoresistivity for reconstruction of the 2D-hole gas (2DHG) spectrum under quantizing magnetic fields in p-Ge/Ge$_{1-x}$Si$_x$ systems with complex valence band structure [7]. Measurements of the longitudinal $\rho_{xx}$ and Hall $\rho_{xy}$ resistivities have been carried out for multilayer p-Ge/Ge$_{1-x}$Si$_x$ ($x = 0.07$) heterostructures with hole concentration $p = (2.4 \div 2.6) \cdot 10^{11}$ cm$^{-2}$ and mobilities $\mu_p = (1.1 \div 1.7) \cdot 10^4$ cm$^2$/Vs in magnetic fields up to 12T at $T = (0.1 \div 15)$K.

The heterostructures consist of 20-nm Ge and Ge$_{1-x}$Si$_x$ layers repeated $15 \div 30$ times. The 2DHG forms inside the undoped Ge layer. The central regions of Ge$_{1-x}$Si$_x$ barriers doped with boron are separated from the Ge layers by 5-nm undoped Ge$_{1-x}$Si$_x$ spacer layers ($d_s$) (Fig.1).

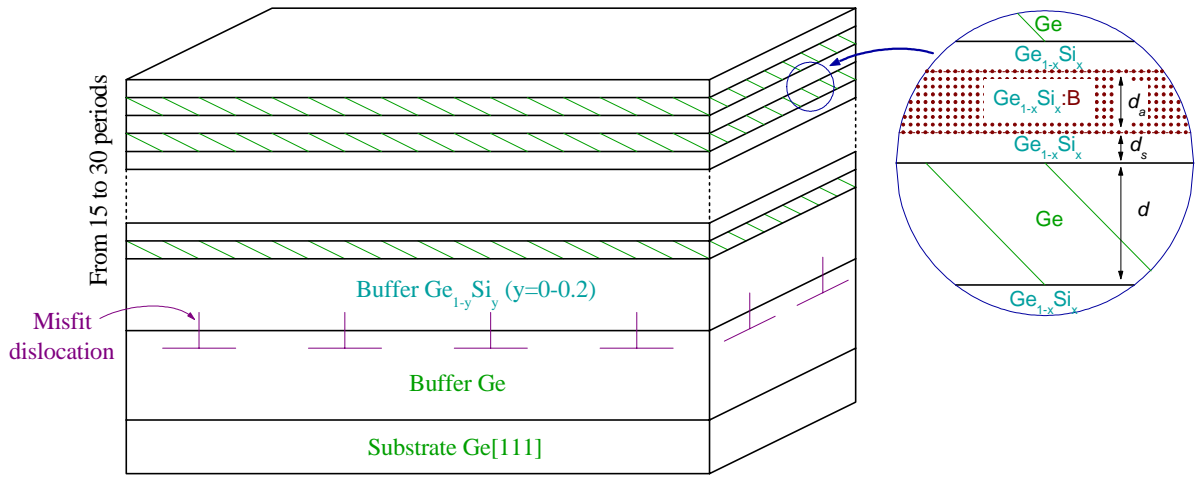

Figure 1.

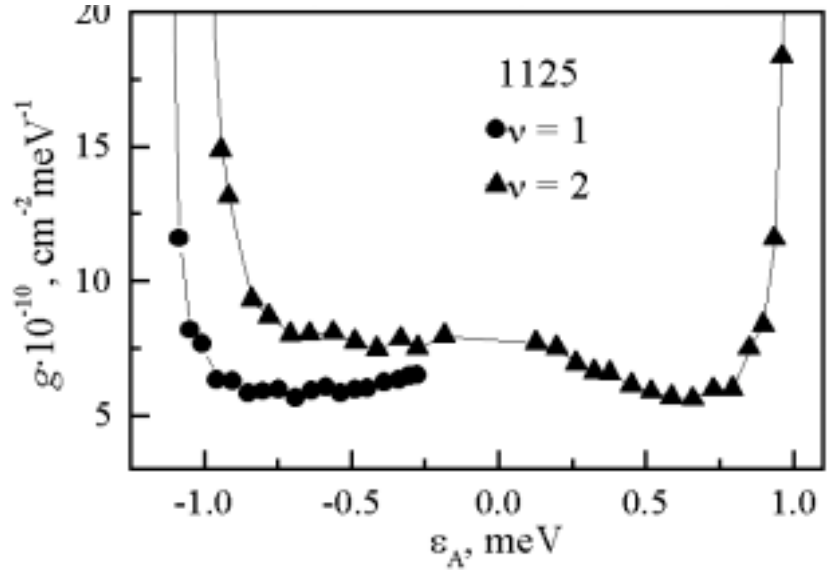

Figure 2.

In the mobility gap DOS obtained as a function of energy $g(\varepsilon)$, even in the middle of a gap when the filling factor is close to an integer, the density of localized states is found to have values comparable with or even higher than the DOS of 2DHG without magnetic field ($g_o \cong 4.5 \cdot 10^{10}$ cm$^{-2}$ meV$^{-1}$). Moreover, $g(\varepsilon)$ remains almost constant in the overwhelming part of the energy intervals between adjacent LL: $g(\varepsilon) \cong g_c = (5 \div 7) \cdot 10^{10}$ cm$^{-2}$ meV$^{-1}$ for $\nu = 1$ and $\nu = 2$ (Fig. 2). This result is consistent qualitatively with the data for structures with *n*-type conductivity [3-6]. As for our value of $g_c$, it is roughly an order of magnitude higher than those for InGaAs/InP [5] and for high-mobility AlGaAs/GaAs [4] heterostructures but comparable with those for Si-MOSFET [3] and intermediate-mobility AlGaAs/GaAs heterostructures [6].

As all short-range impurity potential models lead to an exponential drop in DOS between Landau levels, the clear picture for the DOS in QHE regime may be presented only in terms of the long-range potential fluctuations in combination with the oscillating dependence of DOS on the filling factor. Such an idea has been advanced in early work of Shklovskii and Efros [8] and then developed in series of works of Efros with collaborators (see [9,10] and references therein). In selectively doped heterostructures, the smooth random potential is formed by fluctuations in concentration of remote impurities.

For a random potential $V(r)$ smooth on the scale of magnetic length $l_B$, the localization in QHE regime can be discussed in terms of semiclassical quantization and percolation [11]. In the quasiclassical limit, the electron energy in quantizing magnetic field may be presented as

$$E_N(r_0) = \hbar \omega_c \left(N + \frac{1}{2}\right) + V(r_0) \qquad (1)$$

with $r_0$ being the oscillator center coordinate. Thus the smooth potential removes the degeneracy on $r_0$ and makes the LL energy dependent on spatial coordinates (fig. 3).

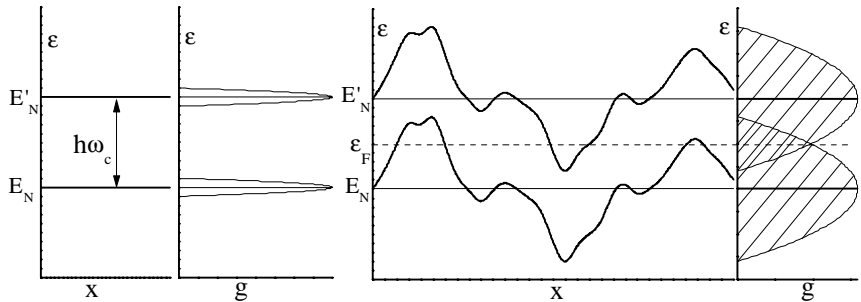

Figure 3.

We report here an order of magnitude evaluation of the spatial scale and amplitude of random potential in p-Ge/Ge$_{1-x}$Si$_x$ heterostructures in QHE regime obtained from an analysis of the mobility gap DOS. Two models for random impurity potential were used.

i) The model with randomly distributed charged centers located within a thick layer close to the 2D-electron (hole) gas [8], for which the relation between fluctuation amplitude $F$ and scale $L$ reads:

$$F(L) = \beta \frac{e^2 \sqrt{NL}}{\kappa}, \qquad (2)$$

$\beta$ is a numerical coefficient ($\beta \cong 0.1$ [9]), $N$ – the density of charged impurities (per volume) and $\kappa$ – the dielectric constant.

ii) The model of the system with a spacer: a condenser with 2D electron (hole) gas as one plate and randomly distributed charged centers as the other plate, separated by a distance $d_s$ [9,10]. In this case:

$$F(L) = \beta \frac{e^2 \sqrt{2\pi C}}{\kappa} \sqrt{\ln \frac{L}{2d_s}}, \qquad (3)$$

where $C$ is the average impurity density (per area).

It is seen from Eqs. (2) and (3) that without screening the amplitude $F$ diverges at large $L$. When the filling factor is close to an integer ($i$) very small concentration of electrons $\delta n \ll n_B$ can be redistributed in space and thus one occurs in conditions of so called nonlinear screening [8-10] ("threshold" screening in terms of [12]). For $\nu = i$ exactly the screening is realized only due to electrons (and holes) induced by an overlap of adjacent fluctuating Landau levels, and so the amplitude of random potential is of the order of corresponding LL gap.

For the investigated heterostructures: $N \cong 10^{17}$ cm$^{-3}$ ($C = Nd_a \cong 10^{11}$ cm$^{-2}$) and the mean distance between impurities $N^{-1/3} \cong 200$Å is comparable both with the width of 2D Ge layer $d \cong 200$Å and the width of doped part of the sample $d_a \cong 100$Å. Thus the described models are not valid precisely but they are suitable to obtain a range of random potential parameter values.

In the nonlinear screening regime, we have the DOS in the middle of mobility gap [8-10] of width $W \cong 2$meV [7] for the two models, respectively:

i) $$g(W/2) = \frac{4\beta e^2 N}{\kappa W^2} \cong 7.5 \cdot 10^{10} \text{ cm}^{-2}\text{meV}^{-1}, \qquad (4)$$

ii) $$g(W/2) = \frac{2\sqrt{C}}{7Wd_s} \cong 9.5 \cdot 10^{10} \text{ cm}^{-2}\text{meV}^{-1}. \qquad (5)$$

So without any fitting parameter we obtain a rather reasonable evaluation of background DOS, and the two models yield values close to each other. For random potential amplitude comparable to the mobility gap, $F \cong W$, we obtain an evaluation of the nonlinear screening length $L_c$ (the scale of optimal fluctuation): $L_c \cong 1000$Å for model (i) (see Eq.(2)) and $L_c \cong 400$Å for model (ii) (see Eq.(3)). As seen in both cases the spatial scale of fluctuations is essentially larger than the magnetic length ($l_B \cong 80$Å at $B = 10$T) hence the random potential may be really regarded as the smooth one.

Thus, an order of magnitude evaluations of the random impurity potential parameters for the p-Ge/Ge$_{1-x}$Si$_x$ heterostructures indicate that in the vicinity of integer filling factors $\nu = 1$ and $\nu = 2$ (i.e. in the regions of the QHE plateaux) a sharp broadening of LL takes place (fig. 3). It is reputed that for the filling factors close to half integers (the regions of plateau to plateau transition) the potential fluctuations would be small due to effective (linear) electron screening [8-10].

**Acknowledgements**

This work is supported in part by RFBR, projects No 99-02-16256, No 01-02-17685 and No 01-02-06131.